\documentclass[aps,prd,12pt,notitlepage,showpacs,nofootinbib,tightenlines]{revtex4-1}
\usepackage{amsmath}
\usepackage{amssymb}
\usepackage{epsfig}
\usepackage{graphicx}
\usepackage{bm}
\usepackage{times}
\usepackage{braket}
\usepackage{color}
\usepackage{slashed}
\usepackage{hyperref}
\definecolor{Red}{rgb}{1.,0.,0.}

\definecolor{Blue}{rgb}{0.,0.,1.}

\definecolor{nicered}{rgb}{0.7,0.1,0.1}
\definecolor{nicegreen}{rgb}{0.1,0.5,0.1}
\hypersetup{colorlinks,citecolor=nicegreen,linkcolor=nicered}

\begin{document}
%%%%%%%%%%%%%%%%%%%%%%%%%%%%%%%%%%%%%%%%%%%%%
\newcommand{\beq}{\begin{eqnarray}}
\newcommand{\eeq}{\end{eqnarray}}

\newcommand{\non}{\nonumber\\ }
\newcommand{\ov}{\overline}

\newcommand{\calo}{ {\cal O}}
\newcommand{\calb}{ {\cal B}}
\newcommand{\calr}{ {\cal R}}

\newcommand{\rmt}{ {\rm T}}
\newcommand{\jpsi}{J/\psi}
\newcommand{\etab}{\bar{\eta} }
\newcommand{\etar}{\eta^\prime }
\newcommand{\etap}{\eta^{(\prime)}}
\newcommand{\psl}{ p \hspace{-2.0truemm}/ }
\newcommand{\qsl}{ q \hspace{-2.0truemm}/ }
\newcommand{\epsl}{ \epsilon \hspace{-2.0truemm}/ }
\newcommand{\nsl}{ n \hspace{-2.2truemm}/ }
\newcommand{\vsl}{ v \hspace{-2.2truemm}/ }

%%%%%%%%%%%%%%%%%%%
\def \appb{{ Acta Phys. Polon. B } }
\def \jpcs{{ J. Phys. Conf. Ser } }
\def \cpc{{ Chin. Phys. C } }
\def \ctp{{ Commun. Theor. Phys. } }
\def \epjc{{ Eur. Phys. J. C} }
\def \jhep{ { JHEP } }
\def \jpg{ { J. Phys. G} }
\def \mpla{ { Mod. Phys. Lett. A } }
\def \npb{ { Nucl. Phys. B} }
\def \plb{ { Phys. Lett. B} }
\def \ppnp{ { Prog. Part. $\&$ Nucl. Phys.} }
\def \pr{ { Phys. Rept. } }
\def \prd{ { Phys. Rev. D} }
\def \prl{ { Phys. Rev. Lett.}  }
\def \ptp{ { Prog. Theor. Phys. }  }
\def \zpc{ { Z. Phys. C}  }
\def \csb{ { Chin. Sci. Bull.}  }
\def \ijmpa{ { Int. J. Mod. Phys. A}  }
%%%%%%%%%%%%%%%%%%%%%%%%%%%%%%%%%%%%%%%%%%%%%%%%%%%%%%%%%%%
\title{The quasi-two-body decays $B_{(s)} \to  (D_{(s)},\bar{D}_{(s)}) \rho \to (D_{(s)}, \bar{D}_{(s)})\pi \pi$
in the perturbative QCD factorization approach }
\author{Ai-Jun Ma$^1$} \email{theoma@163.com}
\author{Ya Li$^1$}\email{liyakelly@163.com}
\author{Wen-Fei Wang$^2$}\email{wfwang@sxu.edu.cn}
\author{Zhen-Jun Xiao$^{1,3}$ } \email{xiaozhenjun@njnu.edu.cn}
\affiliation{$^1$ Department of Physics and Institute of Theoretical Physics,
                          Nanjing Normal University, Nanjing, Jiangsu 210023, P.R. China}
\affiliation{$^2$ Institute of Theoretical Physics, Shanxi University, Taiyuan, Shanxi 030006, P.R. China}
\affiliation{$^3$ Jiangsu Key Laboratory for Numerical Simulation of Large Scale Complex
Systems, Nanjing Normal University, Nanjing, Jiangsu 210023, P.R. China}
\date{\today}
%-----------------------------------------------------%
\begin{abstract}
In this paper, we studied the $B_{(s)} \to  (D_{(s)},\bar{D}_{(s)}) \rho \to (D_{(s)}, \bar{D}_{(s)})\pi \pi$
decays by employing a framework for the quasi-two-body decays in the perturbative QCD (PQCD) factorization approach.
We use the two-pion distribution amplitudes $\Phi_{\pi\pi}$, which contains both resonant and nonresonant contributions
from the pion pair, to describe the final state interactions (FSIs) between the pions in the resonant region.
We found that
(a) for all considered decays, the PQCD predictions for their branching ratios based on the quasi-two-body
and the two-body framework agree well with each other due to ${\cal B}(\rho \to \pi\pi) \approx 100\%$;
For $B^+\to \bar{D^0}\rho^+\to \bar{D^0}\pi^+ \pi^0$ and other four considered decay modes,
the PQCD predictions do agree well with the measured values within errors;
(b) the great difference between the PQCD predictions for ${\cal B}(B\to \bar{D} \rho \to \bar{D}
\pi\pi)$ and ${\cal B}(B\to D \rho \to D \pi\pi)$ can be understood by
the strong CKM suppression factor $ R_{\rm CKM} \approx 3 \times 10^{-4}$;
(c) for the $B_s\to D \rho \to D \pi\pi$ and $B_s\to \bar{D} \rho \to \bar{D} \pi\pi$ decays,
however, the PQCD predictions of $R_{\rm s1}\approx 0.13$ and $R_{\rm s2}\approx 0.14$ do agree very well with
the moderate CKM suppression factor $R^s_{\rm CKM}\approx 0.14$;
and (d) the PQCD predictions for the ratios $R_{D\rho}$ and  the strong phase difference $\cos\delta_{D\rho}$
of the three $B \to \bar{D} \rho$ decay modes agree well with the LHCb measurements within one standard deviation.
\end{abstract}


\maketitle
%------------------------------------------------------%

\section{Introduction}\label{sec:1}

The hadronic two-body and three-body $B$ meson decays provide rich information for studying the heavy flavor physics in
and beyond the Standard Model (SM), but the three-body decays are clearly more complicated than the two-body cases
due to the involvement of the resonant and nonresonant contributions, as well as the possible FSIs.
In experiments, the BaBar~\cite{BaBar,BaBar1,BaBar2,BaBar3,BaBar4},
 Belle~\cite{Belle,Belle1,Belle2,Belle3,Belle4,Belle5} and  LHCb Collaborations~\cite{LHCb,LHCb1,LHCb2,LHCb3,LHCb4,LHCb5}
have reported their measurements for the branching ratios and CP violations of some hadronic three-body $B/B_s$
meson decay modes.
The large localized CP asymmetries in a number of such decay channels, specifically,
have raised great interests in theoretical studies~\cite{Li:2003,Li:2004,wenfei1,wenfei2,Li2,li17a,Cheng,
Cheng1,Cheng2,Cheng3,Cheng4,Cheng5,Furman,Furman1,Furman2,Furman3,
Bhattacharya,Gronau,London,London1,London2,Wang,Lesniak,LiYing,He1,He2,Yang,I,I1,Susanne,wenfei} .

It is fair to say that the nonresonant contributions in  the three-body $B$-meson
decays are quite difficult to calculate,  since we can not separate the nonresonant contributions from the
resonant ones clearly and have no good methods to estimate the non-resonant contributions reliably~\cite{Susanne}.
In the so called ``quasi-two-body" approximation, the two-body scattering and all possible interactions
between the two involved  particles are included but the interactions between the third particle (usually
referred to as bachelor) and the pair of mesons are ignored.

In a recent work~\cite{wenfei}, the authors studied the quasi-two-body decay $B \to K \rho \to K \pi \pi$
by employing the PQCD factorization approach based on $k_T$ factorization theorem.
The resonant and nonresonant contributions between two final pions are parameterized
into the time-like pion form factors involved in the $P$-wave two-pion distribution amplitudes $\phi^{I=1}_{\pi\pi}$,
the PQCD predictions for the branching ratios and the CP-violating asymmetries as presented in Ref.~\cite{wenfei}
are in good agreement with currently available experimental measurements.
By analyzing the distribution of the branching ratios and direct CP asymmetries in the
pion-pair invariant mass $w$, they also found that the main
portion of the branching ratios lies in the region around the pole mass of the $\rho$ resonance as expected.
For $B^+\to K^+\rho^0\to K^+ \pi^+\pi^-$ decay,  for example, its differential decay rate $d{\cal B}/dw$
exhibits peak at $w=m_\rho$, the central value of its branching ratio is ${\cal B}=2.46\times 10^{-6}$
in the range of $ w=[m_\rho -\Gamma_\rho, m_\rho +\Gamma_\rho]$, which is around $72\%$ of the total decay rate
${\cal B}=3.42 \times 10^{-6}$ \cite{wenfei}.
In Ref.~\cite{li17a}, we extend this work \cite{wenfei} to the cases $B\to P\rho\to P\pi\pi$,
where the $P$ standing for kaon and other light pseudoscalar mesons $(\pi, \eta,\eta^\prime)$ as well.
For all $B\to P\rho\to  P \pi\pi$ decays studied in Refs.~\cite{wenfei,li17a},
the PQCD predictions for their branching ratios of those quasi-two-body modes in the three-body and the
two-body frameworks are well consistent with each other. This fact is generally expected since
${\cal B}(\rho \to \pi\pi) \approx 100 \% $, and it does suggest that the PQCD approach is a
consistent theory for exclusive hadronic $B$ meson decays \cite{wenfei,li17a}.

Besides the above mentioned $B$ meson charmless decays, the two-body hadronic charmed decays $B \to D M$ ($M$ denotes
the light pseudoscalar and vector mesons ) have also been
studied  by many authors based on rather different theoretical approaches~
\cite{Beneke,Bauer,Chiang,sihong,B-D1,B-D11,B-D2,B-D21,B-D3,B-D30,B-D31,Lu1,Lu2,Lu11,Lu21}.
Since such charmed hadronic $B$ decay modes involve the tree operators $O_{1,2}$ only, there are much
less theoretical uncertainties from the relevant QCD dynamics.

In the PQCD factorization approach, the factorization for $B \to D M$ decays  was approved at
the leading order of $m_D/m_B$ expansion~\cite{B-D1,B-D11}. Many two-body charmed decays
$B_{(s)} \to  D^{(*)}_{(s)} (P,V,T)$ have been studied in Refs.~\cite{B-D2,B-D21,B-D3,B-D30,B-D31,Lu1,Lu2,Lu11,Lu21}.
In Refs.~\cite{Lu1,Lu11}, specifically, the authors analyzed $B_{(s)} \to  D^{(*)}_{(s)} (P,V)$ decays.
By using the data of six $B \to DP$ channels available at 2008, they firstly made a selection for the expression
of $D/D_s$ meson wave functions by $\chi^2$ fit, presented the PQCD predictions for the considered charmed $B$
decays and found that most of them agreed very well with experiments.
Some predictions for $B_s$ decays \cite{Lu1},
such as ${\cal B}(\bar{B}_s^0 \to D_s^+ K^-)\approx 1.70\times 10^{-4}$ and ${\cal B}(\bar{B}_s^0 \to D_s^{*+} \pi^-)\approx
18.9\times 10^{-4}$, are confirmed \footnote{From HFAG 2016 \cite{hfag2016}, it is easy to find the average
of the measured values: ${\cal B}(\bar{B}_s^0 \to D_s^+ K^-)= (1.92\pm 0.22)\times 10^{-4}$ and
${\cal B}(\bar{B}_s^0 \to D_s^{*+} \pi^-)= (24^{+7}_{-6} ) \times 10^{-4}$. }
by later experimental measurements \cite{pdg2016,hfag2016} .

In this paper, we will extend previous studies as presented in Refs.~\cite{wenfei,li17a} to the cases
of the quasi-two-body charmed decays $B_{(s)} \to  (D_{(s)},\bar{D}_{(s)}) \rho \to (D_{(s)}, \bar{D}_{(s)})\pi \pi$
by employing the PQCD factorization approach, and to examine if the PQCD approach are applicable to the cases
involving a charmed meson as one of the three final state mesons.
Since only the tree diagrams contribute to the considered decay processes, there is no direct CP asymmetry for these
decays in the standard model.
We consider the decays $B_{(s)} \to  \bar{D}_{(s)} \rho \to  \bar{D}_{(s)} \pi \pi$ (through $\bar{b} \to \bar{c}$ transition )
and the CKM suppressed ones $B_{(s)} \to  D_{(s)} \rho \to D_{(s)}\pi \pi$ ( through $\bar{b} \to \bar{u}$ transition),
describe the two-pion system by using the same kind of $P$-wave two-pion distribution amplitudes $\phi^{I=1}_{\pi\pi}$ as in
Refs.~\cite{wenfei,li17a}, present the PQCD predictions for the branching ratios of those considered decays.
By using the same Gegenbauer coefficients for the $\rho$ meson distribution amplitudes, we also do the calculations
in the usual two-body PQCD framework, and compare the numerical results obtained from the different
framework of the two-body decay and the quasi-two-body decay.

This paper is organized as follows. In Sec.~II, we give a brief introduction for the theoretical framework,
calculate and present the decay amplitudes.
The numerical values, some discussions and the conclusions will be given in last two sections.

\section{The theoretical framework}\label{sec:2}

\begin{figure}[]
\begin{center}
\vspace{-3cm} \centerline{\epsfxsize=17cm \epsffile{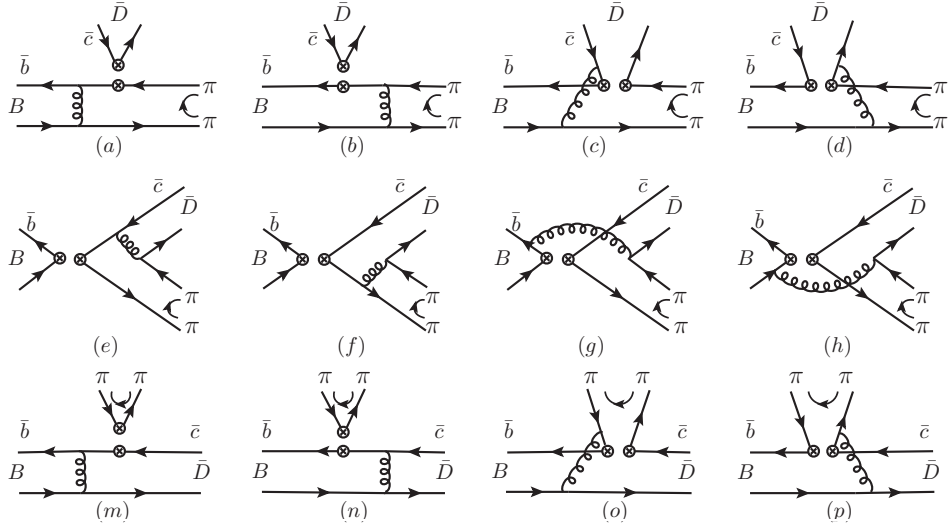}}
\vspace{-14.5cm} \caption{Typical Feynman diagrams for the decays $B_{(s)} \to  \bar{D}_{(s)} \rho \to  \bar{D}_{(s)}
\pi \pi$, where $B_{(s)}= (B^+,B^0,B_s^0)$,
$\bar{D}_{(s)} =(\bar{D^0},D^-,D_s^-)$ and $\rho=(\rho^+, \rho^-, \rho^0)$.}
\label{fig:fig1}
\end{center}
\end{figure}

\begin{figure}[]
\begin{center}
\vspace{-3cm} \centerline{\epsfxsize=16cm \epsffile{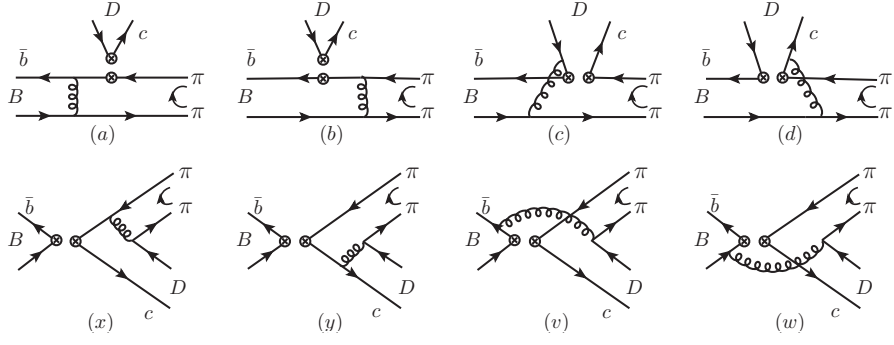}}
\vspace{-16cm} \caption{Typical Feynman diagrams for the CKM-suppressed
decays $B_{(s)} \to  D_{(s)} \rho \to D_{(s)}\pi \pi$, where $ D_{(s)} =(D^0,D^+,D_s^+)$.} \label{fig:fig2}
\end{center}
\end{figure}

For $B_{(s)} \to  \bar{D}_{(s)} \rho \to \bar{D}_{(s)} \pi \pi$ decays and the CKM-suppressed
$B_{(s)} \to  D_{(s)} \rho \to D_{(s)} \pi \pi$ decays, the effective Hamiltomian are of the form
\beq
{\cal  H}_{eff}&=& \left\{\begin{array}{ll}
\frac{G_F}{\sqrt{2}}V^*_{cb}V_{uq}\left[C_1(\mu)O_1(\mu)+C_2(\mu)O_2(\mu)\right],
& \ \  {\rm for} \ \ B_{(s)} \to  \bar{D}_{(s)} \rho \to  \bar{D}_{(s)} \pi \pi\ \ {\rm decays},\\
\frac{G_F}{\sqrt{2}} V^*_{ub}V_{cq}\left[C_1(\mu)O_1(\mu)+C_2(\mu)O_2(\mu)\right],
& \ \  {\rm for} \ \ B_{(s)} \to  D_{(s)} \rho \to D_{(s)}\pi \pi \ \ {\rm decays},\\
\end{array} \right.
\eeq
where $O_{1,2}(\mu)$ are the tree operators, $C_{1,2}(\mu)$ are the Wilson coefficients, $q=(d,s)$ and $V_{ij}$
are the CKM matrix elements. The typical Feynman diagrams for the decays $B_{(s)} \to  \bar{D}_{(s)} \rho \to  \bar{D}_{(s)} \pi \pi$ and $B_{(s)} \to  D_{(s)} \rho \to D_{(s)}\pi \pi$
are shown in the Fig.~\ref{fig:fig1} and \ref{fig:fig2}, respectively.

In the light-cone coordinates, the $B$ meson momentum $p_{B}$, the $\rho$ meson momentum $p$,
and the $D$ meson momentum $p_3$ are chosen as
\beq\label{mom-pBpp3}
p_{B}=\frac{m_{B}}{\sqrt2}(1,1,0_\rmt),~\quad p=\frac{m_{B}}{\sqrt2}(1-r^2,\eta,0_\rmt),~\quad
p_3=\frac{m_{B}}{\sqrt2}(r^2,1-\eta,0_\rmt),
\eeq
where $m_{B}$ denotes the $B$ meson mass, the variable $\eta$ is defined as $\eta=w^2/[(1-r^2)m^2_{B}]$
with the mass ratio  $r=m_{D}/m_{B}$ and the invariant mass squared $w^2=p^2=m^2(\pi\pi)$ of the pion pair.
The momentum of the light quark in the $B$ meson, $\rho$ and $D$ meson are denoted as $k_B$, $k$ and $k_3$ respectively
\begin{eqnarray}
k_B=(0,x_B\frac{m_B}{\sqrt{2}},{k}_{B\rmt}),~\quad
k=(z\frac{(1-r^2)m_B}{\sqrt{2}},0,{k}_{_\rmt}),~\quad
k_3=(0,x_3\frac{(1-\eta)m_B}{\sqrt{2}},{k}_{3_\rmt}),
 \end{eqnarray}
where the momentum fraction $x_{B}$, $z$ and $x_3$ run between zero and unity.
The momentum of pion pair $p_1$, $p_2$ are expressed as
\begin{eqnarray}
p_1^+=\zeta p^+\;,\quad p_2^+=(1-\zeta)p^+\;,\quad p_1^-=(1-\zeta) p^-\;,\quad
p_2^-=\zeta p^-,
\end{eqnarray}
in which $\zeta$ is the momentum fraction for one of the pion pair and $p=p_1+p_2$. For
the decays involving $B_s$ or $D_s$ mesons, one can get the relevant definitions or expressions
from those as given in Eqs.~(2-4) by simple replacement $m_B\to m_{B_s}$ and $m_D\to m_{D_s}$.
All discussions for the $B$ meson decays are applicable for the cases of $B_s$ decays, unless explained
specifically.

The total decay amplitude ${\cal A}$ for the quasi-two-body decays
$B \to D \rho \to D \pi \pi$ in PQCD approach can be written conceptually as the convolution ~\cite{Li:2003,Li:2004,wenfei}
\beq
{\cal A}=\Phi_B\otimes H\otimes \Phi_{D}\otimes\Phi^{I=1}_{\pi\pi},
\eeq
where the hard kernel $H$ describes the dynamics of the strong and electroweak interactions in the decays,
the functions $\Phi_B$, $\Phi_D$ and $\Phi_{\pi\pi}$ are wave functions for the  $B$ meson,
the final state $D/\bar{D}$ meson and the pair of pions, which absorb the non-perturbative
dynamics in the process.

The wave function of $B$ meson can be written  as the one for example in Ref.~\cite{Keum:2000wi}
\beq
\Phi_B= \frac{i}{\sqrt{2N_c}} (\psl_B +m_B) \gamma_5 \phi_B ({\bf k_1}),
\label{bmeson}
\eeq
where the $B$-meson distribution amplitude $ \phi_B $ is of the form
\beq
\phi_B(x,b)&=& N_B x^2(1-x)^2\mathrm{exp} \left
 [ -\frac{M_B^2\ x^2}{2 \omega_{B}^2} -\frac{1}{2} (\omega_{B}\;b)^2\right],
\label{phib}
\eeq
where the normalization factor $N_B$ is determined through the relation
$\int_0^1dx \; \phi_B(x,0)=f_B/(2\sqrt{6})$, $\omega_B$ is a free parameter and one usually
take $\omega_B = 0.40 \pm 0.04$ GeV and $\omega_{B_s}=0.50 \pm 0.05$ GeV
in the numerical calculations ~ \cite{Li:2003,Li:2004,wenfei} for the case of $B$ and $B_s$ decays respectively.

For $D$ meson, in the heavy quark limit, the two-parton LCDAs can be written as \cite{B-D1,B-D11,Lu1,Lu11,Lu2,Lu21}
\beq
\langle D(p_3)|q_{\alpha}(z)\bar{c}_{\beta}(0)|0\rangle
\,&=&\,\frac{i}{\sqrt{2N_{c}}}\int_{0}^{1}dx\,e^{ixp_3\cdot
z}\left[\gamma_{5}(\makebox[-1.5pt][l]{/}p\,_3+\,m_{D})\phi_{D}(x,b)\right]_{\alpha\beta},
\eeq
 where
\beq
\phi_{D}(x,b)=\frac{1}{2\sqrt{2N_{c}}}\,f_{D}\,6x(1-x)\left[1+C_{D}(1-2x)\right]
\exp\left[\frac{-\omega^{2}b^{2}}{2}\right],
\eeq
with $C_{D}=0.5\pm0.1, \omega=0.1$ GeV and $f_{D}=204.6$ MeV for $D$ meson,  and $C_{D_s}=0.4\pm0.1, \omega=0.2$ GeV and
$f_{D_{s}}=257.5$ MeV for $D_{s}$ meson.
In the above models, $x$ is the momentum fraction of the light quark in $D$ ($D_s$) meson.

For the quasi-two-body decays, the dynamics associated with the pair of the pion mesons are factorized into
two-meson distribution amplitudes for the following two reasons~\cite{Li:2003,Li:2004,Pire,npb555-231}:
\begin{enumerate}
\item[(1)] It is not practical to make a direct evaluation for the hard $b$-quark decay kernels containing two virtual
gluons at leading order due to the enormous number of diagrams, while the contribution from
such kinematic region is in fact not important due to the power-suppression;

\item[(2)] The dominant contribution comes from the region where the involved two energetic mesons
are almost collimating to each other and having an invariant mass below $O(\bar\Lambda m_B)$ ($\bar\Lambda=m_B-m_b$).
\end{enumerate}
The longitudinal distribution amplitude of $\rho$ meson is defined as the two-pion distribution
amplitudes~\cite{wenfei}:
\beq
\Phi^{I=1}_{\pi\pi}=\frac{1}{\sqrt{2N_c}}\left[\psl\phi^0(z,\zeta,w^2)
+w\phi^s(z,\zeta,w^2) +\frac{\psl_ 1\psl_2-\psl_2\psl_ 1}{w(2\zeta-1)} \phi^t(z,\zeta,w^2)\right]\; ,
\eeq
where
\begin{eqnarray}
\phi^0(z,\zeta,w^2)&=& \frac{3F_\pi(w^2)}{\sqrt{2N_c}}z(1-z)\left [ 1+a^0_2 C_2^{3/2}(t)\right ]P_1(2\zeta-1),\non
\phi^s(z,\zeta,w^2)&=& \frac{3F_s(w^2)}{2\sqrt{2N_c}}(1-2z)\left [1+a^s_2 (1-10z+10z^2) \right ]P_1(2\zeta-1),\non
\phi^t(z,\zeta,w^2)&=& \frac{3F_t(w^2)}{2\sqrt{2N_c}}(1-2z)^2\left [1+a^t_2 C_2^{3/2}(t) \right ]P_1(2\zeta-1),
\end{eqnarray}
with the Gegenbauer polynomial $C_2^{3/2}(t)=\frac{3}{2}(5t^2-1)$, $t=2z-1$,
the Legendre polynomial $P_1(2\zeta-1)=2\zeta-1$. We choose the Gegenbauer coefficient $a^0_2=0.25\pm0.10$, $a^t_2=
-0.60\pm 0.20$ and $a^s_2=0.75\pm 0.25$ determined in Ref.~\cite{wenfei}.

The strong interactions between the $\rho$ meson and the pion pair with the inclusion of the elastic rescattering
among the two pions  are factorized into $F_\pi$. The form factor $F_\pi$ for $\rho$ meson is chosen as~\cite{JP}
\beq
F_\pi(w^2) &=& \frac{1}{1+\sum c_i}\cdot \left \{ {\rm BW}^{\rm GS}_\rho (w^2, m_\rho, \Gamma_\rho)
\frac{1+c_\omega {\rm BW}^{\rm KS}_\omega(w^2, m_\omega, \Gamma_\omega)}{1+c_\omega}\right. \non
&& \left. \hspace{2cm}+ \sum c_i   {\rm BW}^{\rm GS}_i   (w^2, m_i,  \Gamma_i) \right\}, ~~\label{eq:fpiw2}
\eeq
with
\beq
{\rm BW}^{\rm KS}_{\omega} (s, m, \Gamma)&=& \frac{m^2}{m^2-s-im\Gamma }, \non
{\rm BW}^{\rm GS}_{\rho,i} (s, m, \Gamma)
&=& \frac{m^2\left [ 1+ d(m)\; \Gamma /m \right]}{m^2-s+f(s,m,\Gamma)-im\Gamma(s,m,\Gamma) }, ~~\label{eq:fpiw3}
\eeq
here ${\rm BW}^{\rm KS}_{\omega} (s, m, \Gamma)$ are the ordinary Breit-Wigner (BW) function~\cite{BW} for $\omega$
meson, while  ${\rm BW}^{\rm GS}_{\rho,i}$ are the functions for the $\rho$ and its excited states
$i = (\rho'(1450), \rho''(1700), \rho'''(2254))$ as described by the Gounaris-Sakurai (GS) model~\cite{GJ}.
The explicit expressions of the functions $\Gamma(s,m,\Gamma)$, $d(m)$ and $f(s,m,\Gamma)$ in Eq.~(\ref{eq:fpiw3})
can be found in Eqs.~(29-31) of Ref.~\cite{JP}, other relevant parameters such as $c_\omega$ and $c_i$
in Eq.~(\ref{eq:fpiw2}-\ref{eq:fpiw3}) can also be found in the Appendix of Ref.~\cite{JP}.
Here, we single out the part of $\rho$ meson component.
The equivalence between the framework with the $\rho$ meson
propagator and the present one with the two-pion distribution amplitudes leads to the
relations~\cite{wenfei}:
\beq
F^{\rho}_\pi(w^2) \approx \frac{g_{\rho\pi\pi} w f_\rho}{D_\rho(w^2)},\quad
F^{\rho}_{s,t}(w^2) \approx \frac{g_{\rho\pi\pi} w \; f^T_\rho}{D_\rho(w^2)},~~
\eeq
where $g_{\rho\pi\pi}$ describes the strength of the $\rho \to \pi\pi$ transition, $D_\rho$ represents
the denominator of the BW function for the $\rho$ resonance and $f_\rho(f^T_\rho)$ is associated with
the normalization of the twist-2 ( twist-3) $\rho$ meson distribution amplitudes
( $f_\rho=0.216$ GeV ~\cite{Ball,Bharucha}  and $f^T_\rho=0.184$ GeV~\cite{Jansen}  numerically).

After the integration for $\zeta$, the differential decay rate is written as
\beq
\frac{d{\cal B}}{dw^2}=\tau_{B}\frac{|\vec{p}_\pi|
|\vec{p}_D | }{96\pi^3m^3_{B}}|{\cal A}|^2,
\label{expr-br}
\eeq
where $\tau_{B}$ is the mean lifetime of $B$ meson, and $|\vec{p}_\pi|$ and $|\vec{p}_D|$ denote the magnitudes of
the $\pi$ and $D$ momenta in the center-of-mass frame of the pion pair,
\beq
|\vec{p}_\pi|&=&\frac12\sqrt{w^2-4m^2_{\pi}}, \non
|\vec{p}_D|&=&\frac{1}{2}
\sqrt{[(m^2_B-m^2_D)^2-2(m^2_B+m^2_D)w^2+w^4]/w^2}.
\eeq

For the considered $B_{(s)} \to  \bar{D}_{(s)} \rho \to  \bar{D}_{(s)} \pi \pi$ decays, the analytic formula for
the corresponding decay amplitudes are of the following form:
\beq
\mathcal{A}({B^+ \to {\bar{D^0}} \rho^+ (\rho^+ \to \pi^+\pi^0)})&=&\frac{G_F}{\sqrt2}V^*_{cb}V_{ud} [a_1F_{e\rho}^{LL}+C_2
M_{e\rho}^{LL}+a_2F_{eD}^{LL}+C_1M_{eD}^{LL} ], \\
\mathcal{A}({B^0 \to {D^-} \rho^+ (\rho^+ \to \pi^+\pi^0)})&=&\frac{G_F}{\sqrt2}V^*_{cb}V_{ud} [a_1F_{a\rho}^{LL}+C_2
M_{a\rho}^{LL}+a_2F_{eD}^{LL}+C_1M_{eD}^{LL} ], \\
\mathcal{A}({B^0 \to {\bar{D^0}} \rho^0 (\rho^0 \to \pi^+\pi^-)})&=&\frac{G_F}{2} V^*_{cb}V_{ud}
[a_1(-F_{e\rho}^{LL}+F_{a\rho}^{LL})+C_2(-M_{e\rho}^{LL}+M_{a\rho}^{LL}) ],
\eeq
\beq
\mathcal{A}({B_s^0 \to {D^-} \rho^+ (\rho^+ \to \pi^+\pi^0)})&=&\frac{G_F}{\sqrt2}V^*_{cb}V_{us} [a_1F_{a\rho}^{LL}+C_2
M_{a\rho}^{LL} ], \\
\mathcal{A}({B_s^0 \to {\bar{D^0}} \rho^0 (\rho^0 \to \pi^+\pi^-)})&=&\frac{G_F}{2}V^*_{cb}V_{us} [a_1F_{a\rho}^{LL}+C_2
M_{a\rho}^{LL} ], \\
\mathcal{A}({B_s^0 \to {D_s^-} \rho^+ (\rho^+ \to \pi^+\pi^0)})&=&\frac{G_F}{\sqrt2}V^*_{cb}V_{ud} [a_2F_{eD}^{LL}+C_1
M_{eD}^{LL} ]
\eeq
where the Wilson coefficients $a_1=C_1+ C_2/3$ and $a_2=C_2+ C_1/3$, the individual amplitude $F_{e\rho}^{LL},$
$M_{e\rho}^{LL},$ $F_{eD}^{LL},$ $M_{eD}^{LL}, F_{a\rho}^{LL}$ and $M_{a\rho}^{LL}$ denote the amplitudes from
different sub-diagrams in  Fig.~1.
\beq
F_{e\rho}^{LL}&=&-8\pi C_F m^4_B f_D\int_0^1 dx_B dz
\int_0^{1/ \Lambda} b_B db_B b db \; \phi_B(x_B,b_B)\non
&&\times \big\{\big[ [r^2 ((1-2\eta)(1+z)+(1-(1-\eta)r^2)z )-(1-\eta) (1+z) ]\phi_0(z)-\sqrt{\eta (1-r^2) }\non
&&\times[(1-\eta)(1-2z(1-r^2))(\phi_s(z)+\phi_t(z))+r^2(\phi_s(z)-\phi_t(z))]  \big] E_e(t_a)h_a(x_B,z,b,b_B)\non
&&\times S_t(z)-\big[ (1-r^2 )   [r^2 (x_B-\eta )-(1-\eta) \eta  ] \phi_0(z)+2\sqrt{\eta (1-r^2) }[(1-\eta)(1-r^2)\non
&&-r^2(x_B-\eta)]\phi_s(z)\big] E_e(t_b)h_b(x_B,z,b_B,b)S_t(|x_B-\eta|)\big\},
\label{eq:f01}
\eeq
\beq
M_{e\rho}^{LL}&=&32\pi C_F m^4_B/\sqrt{6} \int_0^1 dx_B dz dx_3
\int_0^{1/ \Lambda} b_B db_B b_3 db_3 \; \phi_B(x_B,b_B)\; \phi_D(x_3,b_3)\non
&&\times \big\{\big[[(r^2(r^2-\eta)-(1-\eta))(x_B-\eta z-(1-\eta)(1-x_3))+r(\eta r-r_c+(\eta r+r_c) \non
&&\times(r^2-\eta))] \phi_0(z)+\sqrt{\eta (1-r^2) }[r^2((1-\eta)x_3+x_B)(\phi_s(z)+\phi_t(z))- (1-\eta)\non
&&\times(1-r^2)z(\phi_s(z)-\phi_t(z))-2r((1-\eta)r -2r_c)\phi_s(z)]  \big] E_n(t_c)h_c(x_B,z,x_3,b_B,b_3)\non
&&-\big[ ( 1-\eta-r^2(1-2\eta))  ((1-r^2)z+(1-\eta)x_3-x_B ) \phi_0(z)+\sqrt{\eta (1-r^2) }\non
&&\times[r^2(x_B-(1-\eta)x_3) (\phi_s(z)-\phi_t(z)) -(1-\eta)(1-r^2)z(\phi_s(z)+\phi_t(z))]\big]\non
&&\times E_n(t_d)h_d(x_B,z,x_3,b_B,b_3) \big\},
\label{eq:m01}
\eeq
\beq
 F_{a\rho}^{LL}&=&-8\pi C_F m^4_B f_B\int_0^1 dx_3 dz
\int_0^{1/ \Lambda} b_3 db_3 b db \; \phi_D(x_3,b_3)\non
&&\times \big\{\big[ [(1-r^2)(1-2r r_c)-\eta(1-2r^2+2rr_c)-(1-\eta)(1-r^2)^2z]\phi_0(z)\non
&&+\sqrt{\eta (1-r^2) }[r_c(1-\eta)
(\phi_s(z)+\phi_t(z))+r(2(1-r^2)z+rr_c)(\phi_s(z)-\phi_t(z)) \non
&& -4r\phi_s(z) ] \big]E_a(t_e)h_e(z,x_3,b,b_3)S_t(z)
+\big[ [(r^2-1 ) ((1-\eta+r^2)\eta+(1-\eta)^2x_3) ] \phi_0(z)\non
&& +2r\sqrt{\eta (1-r^2) }[(1+\eta)(1+x_3)-2\eta x_3-r^2 ] \phi_s(z)\big] E_a(t_f)h_f(z,x_3,b_3,b)\non
&&\times S_t(|\eta(x_3-1)-x_3|)\big\},
\label{eq:f02}
\eeq
\beq
M_{a\rho}^{LL}&=&32\pi C_F m^4_B/\sqrt{6} \int_0^1 dx_B dz dx_3
\int_0^{1/ \Lambda} b_B db_B b db \; \phi_B(x_B,b_B)\; \phi_D(x_3,b_3)\non
&&\times \big\{\big[ r^2(1-r^2)+\eta(1-\eta) +(1-\eta-r^2(r^2-\eta))(x_3(\eta-1)-x_B+\eta (z-2))\big] \phi_0(z)\non
&&+r\sqrt{\eta (1-r^2) }[(x_B-(1-\eta)(1-x_3))(\phi_s(z)-\phi_t(z))-((1-z)r^2+z)(\phi_s(z)+\phi_t(z))  \non
&&+4 \phi_s(z) ] E_n(t_g)h_g(x_B,z,x_3,b,b_B)+[ ( r^2-\eta-1 )(r^2 ((1-\eta)(1-x_3-z)+x_B-\eta) \non
&&- (1-\eta)(1-z)] \phi_0(z)+r\sqrt{\eta (1-r^2) }[((1-\eta)(1-x_3)+x_B)(\phi_s(z)+\phi_t(z))\non
&&+(r^2(1-z)+z)(\phi_s(z)-\phi_t(z))-2\phi_s(z) ]\big] E_n(t_h)h_h(x_B,z,x_3,b,b_B)
\big\}, \label{eq:m02}
\eeq
\beq
F_{eD}^{LL}&=&-8\pi C_F m^4_B F^\rho_\pi(w^2)\int_0^1 dx_B dx_3
\int_0^{1/ \Lambda} b_B db_B b3 db3 \phi_B(x_B,b_B)\phi_D(x_3,b_3)\non
&&\times \big\{ (1+r)  [r^2+(1-\eta)(1-x_3(1-r)(1-\eta-2r)) ]
E_e(t_m)h_m(x_B,x_3,b_3,b_B)\non
&&\times S_t(x_3)+\big[(1-\eta-r^2)((1-2r)(1+r_c)-\eta)+\eta x_B(1-\eta-2r)\big]\non
&& \times E_e(t_n)h_n(x_B,x_3,b_B,b_3)S_t(x_B) \big\}, \label{eq:f03}
\eeq
\beq
 M_{eD}^{LL}&=&32\pi C_F m^4_B/\sqrt{6} \int_0^1 dx_B dz dx_3
\int_0^{1/ \Lambda} b_B db_B b db \phi_B(x_B,b_B)\phi_D(x_3,b_3)\phi_0(z)\non
&&\times \big\{\big[(x_3+z-2)(\eta r(1-\eta r)+r^2(1-r)(1+r-\eta r))+(1-x_B-z)(1-\eta)\non
&&\times(1+\eta-r^2)+(1-r^2)r(r-x_3)+\eta r(r+x_B)\big] E_n(t_o)h_o(x_B,z,x_3,b_B,b)\non
&&-\big[(1-r)((1-\eta)(1+r)-\eta r)((1-r^2)z-x_B)+(1-\eta)x_3(1-\eta+r(r(2\eta\non
&&+r-1)-1)) \big]E_n(t_p)h_p(x_B,z,x_3,b_B,b) \big\}. \label{eq:m03}
\eeq
The hard function $h_i$ with $i=(a,b,c,d,e,f,g,h,m,n,o,p)$ are obtained from the Fourier transformation
of the hard kernels. The explicit expressions of $h_i$ and the hard scales $t_i$ will be given in Appendix.
The six decay amplitudes ($F_{e\rho}^{LL},\cdots, M_{a\rho}^{LL}$ ) as given in
Eqs.~(\ref{eq:f01}-\ref{eq:m03}) are different from those as given in Eqs.~(31-33,38-40) of Ref.~\cite{Lu1}:
the terms proportional to $r^2$, $r \eta$ or higher order factors  are all kept here but neglected in
Ref.~\cite{Lu1}, although the resulted changes in the PQCD predictions for branching ratios  are not
large in magnitude.

The evolution factors $E_e(t)$, $E_a(t)$ and $E_n(t)$ in above equations are written as the form
\beq
E_e(t)&=&\alpha_s(t) \exp[-S_B(t)-S_\rho(t)],\non
E_a(t)&=&\alpha_s(t) \exp[-S_D(t)-S_\rho(t)],\non
E_n(t)&=&\alpha_s(t) \exp[-S_B(t)-S_\rho(t)-S_D(t)],
\eeq
where the Sudakov exponents are defined as
\beq
S_B&=& S(x_B\frac{m_B}{\sqrt2},b_B )+\frac53\int^t_{1/b_B}\frac{d\bar\mu}{\bar\mu} \gamma_q(\alpha_s(\bar\mu)),\\
S_\rho&=& S(z(1-r^2)\frac{m_B}{\sqrt2},b )+ S((1-z)(1-r^2)\frac{m_B}{\sqrt2},b )+ 2\int^t_{1/b}\frac{d\bar\mu}{\bar\mu} \gamma_q(\alpha_s(\bar\mu)),\\
S_D&=& S(x_3(1-\eta)\frac{m_B}{\sqrt2},b_3 ) +2\int^t_{1/b_3}\frac{d\bar\mu}{\bar\mu} \gamma_q(\alpha_s(\bar\mu)),
\eeq
with the quark anomalous dimension $\gamma_q=-\alpha_s/\pi$. The explicit expressions of the
functions $( S(x_B m_B/\sqrt2,b_B ), \cdots)$  can be found for example in Appendix A of Ref. ~\cite{wang}.
The threshold resummation factor $S_t(x)$ in Eqs.~(\ref{eq:f01},\ref{eq:f02}) is of the form ~\cite{wang}:
\beq
\label{eq-def-stx}
S_t(x)=\frac{2^{1+2c}\Gamma(3/2+c)}{\sqrt{\pi}\Gamma(1+c)}[x(1-x)]^c.
\eeq
We here choose $c=0.3$ in numerical calculations.

For the eight CKM suppressed $B_{(s)} \to  D_{(s)} \rho \to D_{(s)}\pi \pi$ decays, on the other hand,
we can find the same set of
analytic formula for the decay amplitudes and relevant functions by following the same procedure as for
$B_{(s)} \to  \bar{D}_{(s)} \rho \to \bar{D}_{(s)}\pi \pi$ decays. The explicit expressions of
all relevant decay amplitudes and functions will be given in Appendix.

%%%%%%%%%%%%%%%%%%%%%%%%%%%%%%%%%%%%%%%%%%%%%%%%%%%%

\section{Numerical results}\label{sec:3}

Besides the quantities specified in previous sections, the following input parameters
(the masses, decay constants and QCD scale are in units of GeV) will be used in the numerical calculations \cite{pdg2016}:
\beq
\Lambda^{(f=4)}_{ \overline{MS} }&=&0.25, \quad m_B=5.280, \quad m_{B_s}=5.367,
\quad m_{D^\pm}=1.870,\quad m_{D^0/\bar{D^0}}=1.865, \non
m_{D_s^\pm}&=&1.968, \quad m_{\rho}=0.775, \quad m_{\pi^\pm}=0.140, \quad m_{\pi^0}=0.135, \quad
m_{b}=4.8, \quad m_c=1.27,  \non
f_B&=& 0.19, \quad f_{B_s}=0.236, \quad \tau_{B^0}=1.520\; {\rm ps}, \quad \tau_{B^+}=1.638\; {\rm ps},
\quad\tau_{B_{s}}=1.510\; {\rm ps}.
\label{eq:inputs}
\eeq
For the Wolfenstein parameters $(A,\lambda,\bar{\rho},\bar{\eta})$ we use the following values:
$A=0.811 \pm 0.026,~\lambda=0.22506\pm 0.00050$,~$\bar{\rho} = 0.124_{-0.018}^{+0.019},~\bar{\eta}= 0.356\pm 0.011$.

We calculate the branching ratios of $B_{(s)} \to  (D_{(s)},\bar{D}_{(s)}) \rho \to (D_{(s)}, \bar{D}_{(s)})\pi \pi$ decays in
the quasi-two-body and the two-body framework in the PQCD factorization approach by using the same set of the
Gegenbauer moments.  Taking $B^+ \to \bar{D}^0 \rho^+ \to \bar{D}^0 \pi^+ \pi^0$
as one example, we find the PQCD prediction for its branching ratio in the quasi-two-body framework:
\beq
{\cal B}(B^+\to  \bar{D}^0 \rho^+ \to  \bar{D}^0 \pi^+\pi^0) =
\left [ 115 ^{+58}_{-36}(\omega_B)^{+7}_{-8}(a_\rho) \pm 8(C_D) \right ] \times 10^{-4},
\quad \label{eq:br01a}
\eeq
where the first error comes from  the uncertainties of the input parameters $\omega_B = 0.40 \pm 0.04$ or $\omega_{B_s}=
0.50 \pm 0.05$; the second one are induced by the uncertainties of the Gegenbauer moments:
$a^t_2= -0.60\pm0.20$, $a^0_2=0.25\pm0.10$ and $a^s_2=0.75\pm0.25$; the third one is due to
$C_{D}=0.5\pm 0.1$ or $ C_{D_s}=0.4\pm0.1$.

\begin{table}[h]
\begin{center}
\caption{The PQCD predictions for the branching ratios (in units of $10^{-4}$) of
$ B_{(s)}\to \bar{D}_{(s)}\rho \to \bar{D}_{(s)}\pi\pi$ decays
in the quasi-two-body (second column) and the two-body (third column) framework.
We also list those currently available measured values \cite{pdg2016,hfag2016}  of the two-body cases
and the central values of the theoretical predictions as given in Ref.~\cite{Lu1} and Ref.~\cite{sihong}.}
\label{dpda}
\begin{tabular}{l| l |l |l| l| l  } \hline\hline
{\rm Decays} & {\rm Quasi-two-body} & {\rm Two-body} & {\rm  Data \cite{pdg2016,hfag2016}} & {\rm Two-body}\cite{Lu1} &{\rm FAT \cite{sihong}} \\ \hline
 $ {\cal B}(B^+\to \bar{D^0}\rho^+\to \bar{D^0}\pi^+ \pi^0)$ &  $115 _{-38}^{+59} $ & $116 _{-37}^{+56}$ &  $134 \pm 18$ & $111$ &$105$ \\
 $ {\cal B}(B^0\to  D^-\rho^+ \to D^-\pi^+ \pi^0 )$ &  $82.3_{-29.0}^{+49.2} $ & $ 88.2_{-30.7}^{+49.7} $ &  $79 \pm 13 $& $67.0$ & $ 65.3$\\
 $ {\cal B}(B^0\to \bar{D^0}\rho^0 \to \bar{D^0}\pi^+\pi^-)$ & $1.39_{-0.90}^{+1.24} $ &  $ 1.23_{-0.64}^{+0.90}$ &  $ 2.9 \pm 1.1   $ & $1.99$ & $ 2.60 $ \\ \hline
 $ {\cal B}(B_s^0\to \bar{D^0}\rho^0 \to \bar{D^0}\pi^+\pi^-)$ &  $0.026_{-0.006}^{+0.010} $ &  $0.022_{-0.005}^{+0.006}$ &  $- $ &$0.042$& $0.010 $ \\
 $ {\cal B}(B_s^0\to D^-\rho^+ \to D^-\pi^+\pi^0)$ & $ 0.051_{-0.014}^{+0.022}$ &  $ 0.044_{-0.011}^{+0.012}$ &  $-$ &$0.079$& $0.019 $  \\ \hline
 $ {\cal B}(B_s^0\to D_s^- \rho^+\to D_s^-\pi^+\pi^0)$ &  $77.2_{-25.6}^{+40.2} $ & $79.5_{-26.3}^{+40.6}$ &  $85 \pm 21 $ &$47.0$ &$78.6 $ \\ \hline\hline
\end{tabular}
\end{center}
\end{table}

\begin{table}[h]
\begin{center}
\caption{The PQCD predictions for the branching ratios of the CKM suppressed
$ B_{(s)}\to D_{(s)}\rho \to D_{(s)}\pi\pi$ decays
in the quasi-two-body (second column) and the two-body (third column) framework.
We also list those currently available measured values \cite{pdg2016,hfag2016} of the two-body cases
and the central values of the theoretical predictions as given in Ref.~\cite{Lu11} and Ref.~\cite{sihong}.}
\label{dpdb}
\begin{tabular}{l |l |l| l| l |l } \hline\hline
{\rm Decays} & {\rm Quasi-two-body} & {\rm Two-body} & {\rm  Data }\cite{pdg2016,hfag2016} &{\rm Two-body}\cite{Lu11} & {\rm FAT \cite{sihong}} \\ \hline
$ {\cal B}(B^+\to D^0\rho^+\to D^0\pi^+ \pi^0)(10^{-7}) $ &  $0.50 _{-0.14}^{+0.22}$ & $0.53 _{-0.14}^{+0.26}$ &  $-$ &$0.93$&  $ 4.80 $\\
$ {\cal B}( B^0\to D^+\rho^-\to D^+\pi^-\pi^0)(10^{-7}) $ & $ 7.63_{-3.08}^{+5.92} $ & $9.45 _{-4.89}^{+6.48}$ &  $- $ &$12.7$&  $  9.40 $\\
$ {\cal B}(B^0\to D^0\rho^0\to D^0\pi^+\pi^-)(10^{-7}) $ &  $ 0.13 _{-0.08}^{+0.09} $ & $0.13_{-0.05}^{+0.10}$ &  $- $&$0.34$&  $ 1.20 $ \\
$ {\cal B}(B^+\to D^+\rho^0\to D^+\pi^+\pi^-)(10^{-7}) $ &  $ 5.33 _{-2.65}^{+3.60}$ & $5.99 _{-2.91}^{+3.93}$ &  $- $&$7.50$&  $ 3.30 $ \\ \hline
$ {\cal B}(B_s^0\to D^0\rho^0\to D^0\pi^+\pi^-)(10^{-7})$ &  $3.41 _{-0.75}^{+1.03}$ & $3.13 _{-0.64}^{+0.98} $ &  $- $ &$1.90$&  $ 1.30 $\\
$ {\cal B}(B_s^0\to D^+\rho^-\to D^+\pi^-\pi^0)(10^{-7}) $ & $6.88 _{-1.58}^{+1.98}$ & $6.30 _{-1.29}^{+1.96} $ &  $- $ &$3.70$&  $  2.50 $\\ \hline
$ {\cal B}(B^+\to D_s^+\rho^0\to D_s^+\pi^+\pi^-)(10^{-5}) $ & $1.52 _{-0.82}^{+1.11} $ & $1.82 _{-0.91}^{+1.19}$ &  $ <30$ &$1.94$&  $  1.68$\\
$ {\cal B}(B^0\to D_s^+\rho^-\to D_s^+\pi^-\pi^0)(10^{-5}) $ & $2.82 _{-1.53}^{+2.04} $ & $3.37 _{-1.63}^{+2.19}$ &  $1.1\pm 0.9$ &$3.59$&  $3.12 $ \\ \hline \hline
\end{tabular} \end{center}
\end{table}

The numerical results for all fourteen considered decay modes are listed in Table \ref{dpda} and \ref{dpdb},
where the individual errors have been added in quadrature.
As a comparison, we list the PQCD predictions for the branching ratios in both the quasi-two-body framework and
the ordinary two-body framework \footnote{ We calculated the PQCD predictions in the ordinary two-body framework
as listed in the third column of Table \ref{dpda} and \ref{dpdb}. These numerical results are obtained by using
the formulae as given in Refs.~\cite{Lu1,Lu11} but with the updated Gegenbauer moments and other input parameters,
and they agree well with those as given in Refs.~\cite{Lu1,Lu11} (fifth column). }.
The central values of the theoretical predictions obtained by employing the factorization-assisted
topological-amplitude (FAT) approach \cite{sihong} also be listed in the last column of the two Tables.
For some decay modes considered here, the currently available experimental measurements of the two-body cases
$B_{(s)} \to (\bar{D}_{(s)}, D_{(s)}) \rho$ as given in PDG 2016 \cite{pdg2016} or HFAG 2016 \cite{hfag2016}
are included in Table \ref{dpda} and \ref{dpdb} as well.

\begin{figure}[tb]
\begin{center}
\vspace{-1cm}
\centerline{\epsfxsize=12cm \epsffile{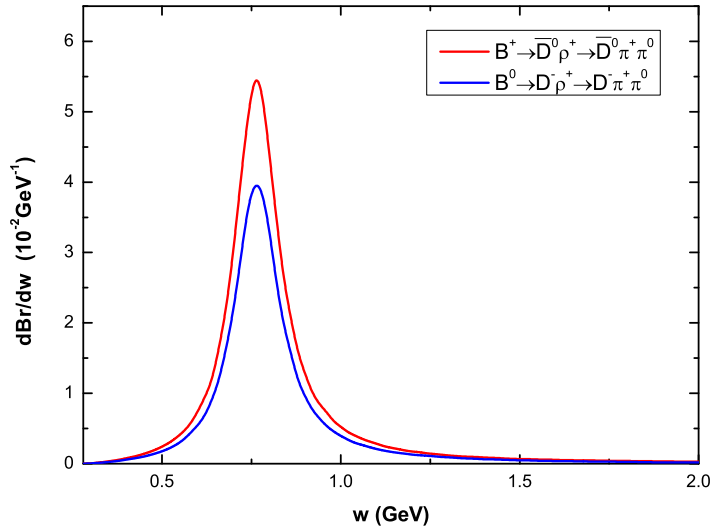} }
\caption{The PQCD predictions for the $w$-dependence of the differential branching ratios for the
$B^+\to \bar{D^0}\rho^+\to \bar{D^0}\pi^+ \pi^0$ (red curve) and
$B^0\to  D^-\rho^+ \to D^-\pi^+ \pi^0$ (blue curve).}\label{fig3}
\end{center}
\end{figure}

Since only the tree diagrams contribute to the considered processes here, there is no direct CP asymmetry for these
considered decays.
From the calculation and numerical results as listed in Table \ref{dpda} and \ref{dpdb},
we have the following observations:
\begin{itemize}
\item[(1)]
Although the PQCD formalism are rather different for the case of the quasi-two-body and the two-body decay analysis,
the PQCD predictions for the branching ratios of all considered decays obtained in both frameworks
do agree very well with each other, as generally expected. The reason in indeed very simple:
${\cal B}(\rho \to \pi\pi ) \approx 100\%$. Consequently, there exist a simple relation between
the decay rate of the same kinds of decays evaluated in the quasi-two-body and the ordinary two-body framework:
\beq
{\cal B}(B_{(s)}\to \bar{D}_{(s)}\rho \to \bar{D}_{(s)}\pi\pi)&=& {\cal B}(B_{(s)}\to \bar{D}_{(s)}\rho) \cdot
{\cal B}(\rho\to \pi\pi) \approx {\cal B}(B_{(s)}\to \bar{D}_{(s)}\rho),\quad  \non
{\cal B}(B_{(s)}\to D_{(s)}\rho \to D_{(s)}\pi\pi)&=& {\cal B}(B_{(s)}\to D_{(s)}\rho) \cdot
{\cal B}(\rho\to \pi\pi) \approx {\cal B}(B_{(s)}\to D_{(s)}\rho).
\eeq
For $B^+\to \bar{D^0}\rho^+\to \bar{D^0}\pi^+ \pi^0$ and other four considered decay modes,
furthermore, the PQCD predictions do agree well with those currently available experimental measurements
\cite{pdg2016,hfag2016} within errors.
We can take above two "good behavior" as a new indication for the reliability of the PQCD factorization approach
and its applicability for the charmed two-body and/or quasi-two-body hadronic decays of $B$ and $B_s$ mesons.

\item[(2)]
For the four CKM suppressed $B\to D \rho \to D \pi\pi$ decays as listed in the first four lines of Table \ref{dpdb},
the PQCD predictions for their branching ratios are much smaller than those for
the three $B\to \bar{D} \rho\to \bar{D} \pi\pi$ decays as given in Table \ref{dpda},
say by about $3-5$ orders. The major reason is the strong CKM suppression factor:
\beq
R_{\rm CKM} = \left | \frac{V_{ub}^* V_{cd}}{V_{cb}^*V_{ud}}\right|^2 \approx \lambda^4 (\bar{\rho}^2 + \bar{\eta}^2)
\approx 3\times 10^{-4},
\eeq
which can be seen easily from the decay amplitudes as given in Eqs.~(17-19) and Eqs.~(A1-A4).
Taking the corresponding pairs of $ B\to \bar{D} \rho$ and  $ B \to D \rho$ decays into account, for example,
the ratios of their branching ratios are of the form
\beq
R_1 &=& \frac{ {\cal B}(B^+\to D^0 \rho^+ \to D^0\pi^+\pi^0)}{{\cal B}(B^+\to\bar{D}^0 \rho^+ \to \bar{D}^0 \pi^+\pi^0)}
\approx 0.44\times 10^{-5}, \non
R_2 &=& \frac{ {\cal B}(B^0\to D^+ \rho^- \to D^+\pi^-\pi^0)}{{\cal B}(B^0\to D^- \rho^+ \to D^- \pi^+\pi^0)}
\approx 0.93 \times 10^{-4}, \non
R_3 &=& \frac{ {\cal B}(B^0\to D^0 \rho^0 \to D^0 \pi^+\pi^-)}{{\cal B}(B^0\to \bar{D}^0 \rho^0 \to \bar{D}^0 \pi^+\pi^-)}
\approx 0.94\times 10^{-4}.
\eeq

\item[(3)]
For the $B_s\to D \rho \to D \pi\pi$ decays and $B_s\to \bar{D} \rho \to \bar{D} \pi\pi$ decays,
there still exist the CKM suppression but it is now much moderate in size than the  previous cases:
\beq
R^s_{\rm CKM} = \left | \frac{V_{ub}^* V_{cs}}{V_{cb}^*V_{us}}\right|^2
\approx (\bar{\rho}^2 + \bar{\eta}^2)\approx 0.14.
\eeq
We can again define the ratios of the branching ratios for the corresponding pairs of $B_s $ decays in the following form:
\beq
R_{\rm s1} &=& \frac{ {\cal B}(B_s^0\to D^0 \rho^0 \to D^0\pi^+\pi^-)}{{\cal B}(B_s^0\to \bar{D}^0 \rho^0 \to \bar{D}^0 \pi^+\pi^-)}
\approx 0.13, \non
R_{\rm s2} &=& \frac{ {\cal B}(B_s^0\to D^+ \rho^- \to D^+\pi^-\pi^0)}{{\cal B}(B_s^0\to D^- \rho^+ \to D^- \pi^+\pi^0)}
\approx 0.14.
\eeq
The PQCD predictions for both $R_{\rm s1}$ and $R_{\rm s2}$ indeed agree very well with $R^s_{\rm CKM}$.

\item[(4)]
In Fig.~\ref{fig3}, we show the $w$-dependence of the differential decay rate $d{\cal B}/dw$ for the first two
decay modes listed in Table \ref{dpda}.
One can see directly that the main  contribution to the decay rates lies in the region around the pole mass
$m_\rho=775$ MeV of the $\rho$ resonance.
Taking the decay $B^+\to \bar{D^0}\rho^+\to \bar{D^0}\pi^+ \pi^0$ as an example, the central values (in units of $10^{-4}$)
of its branching  ratios after making the integration over different ranges of $w$ are of the following form:
\beq
{\cal B}(B^+\to \bar{D^0}\rho^+\to \bar{D^0}\pi^+ \pi^0)&=&\left\{\begin{array}{ll}
89,  & {\rm for} \quad w=[m_\rho-\Gamma_\rho, m_\rho+\Gamma_\rho],\\
109,  & {\rm for} \quad w=[m_\rho-3\Gamma_\rho, m_\rho+3\Gamma_\rho],\\
115, & {\rm for} \quad 2m_\pi \leq w\leq m_B-m_D.\\
\end{array} \right.
\eeq
This is an indication that the quasi-two-body framework is a very good approximation for the charmed $B/B_s$ three-body
decays considered in this paper.

\item[(5)]
The color-allowed emission diagrams Fig.\ref{fig:fig1}(a), \ref{fig:fig1}(b), \ref{fig:fig2}(a) and \ref{fig:fig2}(b)
are generally dominant for the considered decays,
but the color-suppressed nonfactorizable emission diagrams  and annihilation diagrams can also provide
considerable contributions to those decays with the $D$ or $D_s$ meson as one of the final state mesons.

\end{itemize}

Because of the isospin symmetry \cite{LHCb5}, there is a relation between the decay amplitudes $A_{1/2}$ and
$A_{3/2}$ of the charmed decays $B^+ \to \bar{D^0} \rho^+$, $B^0 \to D^- \rho^+$ and
$B^0 \to \bar{D^0} \rho^0$ considered here:
\beq
A(\bar{D^0}\rho^+)=A(D^-\rho^+)+\sqrt{2}A(\bar{D^0}\rho^0).
\eeq
Based on such isospin symmetry, one can further define the amplitude ratio $R_{D\rho}$ and
the strong phase difference $\delta_{D\rho}$ between the amplitudes $A_{1/2}$ and $A_{3/2}$ in the following forms~\cite{LHCb5}
\beq
R_{D\rho} &=& \sqrt{\frac{1}{2}}\sqrt{ \frac{\tau_{B^+}}{\tau_{B^0}}\cdot
\frac{3\,({\cal B}(D^-\rho^+) + {\cal B}(\bar{D^0} \rho^0))}{{\cal B}(\bar{D^0} \rho^+)} - 1 }, \\
\cos\delta_{D\rho} &=& \frac{1}{4R_{D\rho}} \times \left[ \frac{\tau_{B^+}}{\tau_{B^0}}\cdot
\frac{3\,({\cal B}(D^-\rho^+) - 2\,{\cal B}(\bar{D^0}\rho^0))}{{\cal B}(\bar{D^0}\rho^+)} + 1  \right].
\eeq
We found $R_{D\rho}=0.82$ $(0.86)$ and $\cos\delta_{D\rho}=0.983$ $ (0.981)$ in the quasi-two-body (two-body)
PQCD framework,
while the measured values from LHCb Collaboration ~\cite{LHCb5} are of the form:
\beq
R_{D\rho}&=&0.69\pm0.15, \quad \cos\delta_{D\rho}=0.984_{-0.048}^{+0.113}, \quad {\rm for\ \  Isobar \ \  model}, \\
R_{D\rho}&=&0.69\pm0.15, \quad \cos\delta_{D\rho}=0.987_{-0.048}^{+0.114}, \quad {\rm for \ \ K-matrix \ \ model.}
\eeq
It is easy to see that our PQCD predictions for both $R_{D\rho}$ and  $\cos\delta_{D\rho}$ agree well with
the measured values within one standard deviation.

\section{Summary}\label{sec:4}

In this paper, we studied the contributions from the $P$-wave resonant
states $\rho$ to the   $B_{(s)} \to  (D_{(s)},\bar{D}_{(s)}) \rho \to (D_{(s)}, \bar{D}_{(s)})\pi \pi$
decays.
We considered fourteen decay modes, calculated the branching ratios by employing the quasi-two-body and the two-body
framework respectively in the PQCD factorization approach.
The two-pion distribution amplitude $\Phi^{I=1}_{\pi\pi}$ was introduced, the time-like form factor $F_\pi$ was
employed to describe the strong interactions between the resonance $\rho$ and the pion pair including two meson
final state interactions.

From the analytical and numerical calculations, we found the following points:
\begin{itemize}
\item[(1)]
For all considered decays, the PQCD predictions based on the quasi-two-body and the two-body framework
agree well with each other, as generally expected.
For $B^+\to \bar{D^0}\rho^+\to \bar{D^0}\pi^+ \pi^0$ and other four considered decay modes,
the PQCD predictions do agree well with the measured values within errors.
It is a new indication for the reliability of the PQCD factorization approach
and its applicability for the charmed hadronic decays of $B$ and $B_s$ mesons.

\item[(2)]
The great difference between the PQCD predictions for the branching ratios of $B\to \bar{D} \rho \to \bar{D}
\pi\pi$ decays and the $B\to D \rho \to D \pi\pi$ decays can be explained by
a strong CKM suppression factor $ R_{\rm CKM} \approx \lambda^4 (\bar{\rho}^2 + \bar{\eta}^2) \approx 3 \times 10^{-4}$.

\item[(3)]
For the $B_s\to D \rho \to D \pi\pi$ and $B_s\to \bar{D} \rho \to \bar{D} \pi\pi$ decays,
however, the CKM suppression factor is moderate in size: $R^s_{\rm CKM}\approx (\bar{\rho}^2 + \bar{\eta}^2)
\approx 0.14$, which agrees very well with the PQCD prediction $R_{\rm s1}\approx 0.13$ and $R_{\rm s2}\approx 0.14$.

\item[(4)]
The PQCD predictions for the ratios $R_{D\rho}$ and  the strong phase difference $\cos\delta_{D\rho}$
defined based on the isospin symmetry between the three $B \to \bar{D} \rho$ decay modes agree well with the
LHCb measurements.

\end{itemize}

\begin{acknowledgments}

Many thanks to Hsiang-nan Li, Cai-Dian L\"u and Xin Liu for valuable discussions.
This work was supported by the National Natural Science
Foundation of China under the No.~11235005 and 11547038.

\end{acknowledgments}

\appendix

\section{Some decay amplitudes and relevant functions}

The analytic formula for the $B_{(s)} \to  D_{(s)} \rho \to D_{(s)}\pi \pi$ decay amplitudes are of the following form:
\beq
\mathcal{A}({B^+ \to {D^0} \rho^+ (\rho^+ \to \pi^+\pi^0)})&=&\frac{G_F}{\sqrt2}V^*_{ub}V_{cd}
[a_1F_{e\rho}^{LL}+C_2M_{e\rho}^{LL}+a_2F_{aD}^{LL}+C_1M_{aD}^{LL} ], \\
\mathcal{A}({B^+ \to {D^+} \rho^0 (\rho^0 \to \pi^+\pi^-)})&=&\frac{G_F}{2} V^*_{ub}V_{cd} [a_2(F_{e\rho}^{LL}-F_{aD}^{LL})
+C_1(M_{e\rho}^{LL}-M_{aD}^{LL})], \\
\mathcal{A}({B^0 \to {D^0} \rho^0 (\rho^0 \to \pi^+\pi^-)})&=&\frac{G_F}{2} V^*_{ub}V_{cd}
[a_1(-F_{e\rho}^{LL}+F_{aD}^{LL})+C_2(-M_{e\rho}^{LL}+M_{aD}^{LL}) ],\\
\mathcal{A}({B^0 \to {D^+} \rho^- (\rho^- \to \pi^-\pi^0)})&=&\frac{G_F}{\sqrt2}V^*_{ub}V_{cd} [a_2F_{e\rho}^{LL}+C_1
M_{e\rho}^{LL}+a_1F_{aD}^{LL}+C_2M_{aD}^{LL} ],
\eeq
\beq
\mathcal{A}({B^+ \to {D_s^+} \rho^0 (\rho^0 \to \pi^+\pi^-)})&=&
\frac{G_F}{2}V^*_{ub}V_{cs} [a_2F_{e\rho}^{LL}+C_1M_{e\rho}^{LL} ],\\
\mathcal{A}({B^0 \to {D_s^+} \rho^- (\rho^- \to \pi^-\pi^0)})&=&\frac{G_F}{\sqrt2}V^*_{ub}V_{cs}
[a_2F_{e\rho}^{LL}+C_1M_{e\rho}^{LL} ],\\
\mathcal{A}({B_s^0 \to {D^0} \rho^0 (\rho^0 \to \pi^+\pi^-)})&=&\frac{G_F}{2}V^*_{ub}V_{cs} [a_1F_{aD}^{LL}+C_2M_{aD}^{LL} ],\\
\mathcal{A}({B_s^0 \to {D^+} \rho^- (\rho^- \to \pi^-\pi^0)})&=&\frac{G_F}{\sqrt2}V^*_{ub}V_{cs}
[a_1F_{aD}^{LL}+C_2M_{aD}^{LL} ],
\eeq
with the functions $F_{e\rho}^{LL}, M_{e\rho}^{LL}, F_{aD}^{LL}$ and $M_{aD}^{LL}$ denote the amplitudes  as illustrated by Fig.~2.
\beq
F_{e\rho}^{LL}&=&-8\pi C_F m^4_B f_D\int_0^1 dx_B dz
\int_0^{1/ \Lambda} b_B db_B b db \; \phi_B(x_B,b_B)\non
&&\times \big\{\big[ [r^2 ((1-2\eta)(1+z)+(1-(1-\eta)r^2)z )-(1-\eta) (1+z) ]\phi_0(z)-\sqrt{\eta (1-r^2) }\non
&&\times[(1-\eta)(1-2z(1-r^2))(\phi_s(z)+\phi_t(z))+r^2(\phi_s(z)-\phi_t(z))]  \big] E_e(t_a)h_a(x_B,z,b,b_B)\non
&&\times S_t(z)-\big[ (1-r^2 )   [r^2 (x_B-\eta )-(1-\eta) \eta  ] \phi_0(z)+2\sqrt{\eta (1-r^2) }[(1-\eta)(1-r^2)\non
&&-r^2(x_B-\eta)]\phi_s(z)\big] E_e(t_b)h_b(x_B,z,b_B,b)S_t(|x_B-\eta|)\big\},
\eeq
\beq
M_{e\rho}^{LL}&=&-32\pi C_F m^4_B/\sqrt{6} \int_0^1 dx_B dz dx_3
\int_0^{1/ \Lambda} b_B db_B b_3 db_3 \phi_B(x_B,b_B)\phi_D(x_3,b_3)\non
&&\times \big\{\big[[(1-\eta +r^2) (1 -r^2)((\eta-1)x_3+x_B-\eta z)] \phi_0(z)+\sqrt{\eta (1-r^2) }[r^2((1-\eta)x_3\non
&&-x_B)(\phi_s(z)+\phi_t(z))+(1-\eta)(1-r^2)z(\phi_s(z)-\phi_t(z))  ]  \big] E_n(t_c)h_c(x_B,z,x_3,b_B,b_3)\non
&&-\big[[rr_c(1+\eta)-r^2(r^2(1-2\eta)(z-1)+rr_c-2\eta^2(1-x_3)+\eta(3(z-x_3)+2(1-x_B))\non
&&+x_B+x_3-2z)-(1-\eta)((1-\eta)(1-x_3)-x_B+z)\big]    \phi_0(z)-\sqrt{\eta (1-r^2) }\big[ (\eta-1)\non
&&\times(1-r^2)z(\phi_s(z)+\phi_t(z))+r^2((1-\eta)x_3+x_B)(\phi_s(z)-\phi_t(z))+2 (2rr_c-(1-\eta)r^2)\non
&&\times\phi_s(z)]\big] E_n(t_d)h_d(x_B,z,x_3,b_B,b_3) \big\},
\eeq
\beq
 F_{aD}^{LL}&=&-8\pi C_F m^4_B f_B\int_0^1 dx_3 dz
\int_0^{1/ \Lambda} b_3 db_3 b db \phi_D(x_3,b_3)\non
&&\times \big\{\big[  (1-r^2)[\eta(1-\eta+r^2)+(1-\eta)^2 x_3]\phi_0(z)+2 r\sqrt{\eta (1-r^2) }[  1+\eta+(1-\eta)x_3\non
&&-r^2] \phi_s(z) \big] E_a(t_y)h_y(z,x_3,b_3,b)S_t(x_3)-[ (1-\eta)(r^4(z-1)+r^2(1+\eta-2z)+z)\non
&&+2rr_c(r^2-1-\eta)] \phi_0(z)+\sqrt{\eta (1-r^2) } [r(2z+2r^2(1-z)-rr_c)(\phi_s(z)+\phi_t(z))\non
&&+(1-\eta)(2r-r_c)(\phi_s(z)-\phi_t(z))]\big] E_a(t_y)h_y(z,x_3,b,b_3)S_t(z) \big\}, \\
 M_{aD}^{LL}&=&32\pi C_F m^4_B/\sqrt{6} \int_0^1 dx_B dz dx_3
\int_0^{1/ \Lambda} b_B db_B b db \phi_B(x_B,b_B)\phi_D(x_3,b_3)\non
&&\times \big\{\big[  [r^2(r^2((1-\eta)(x_3-z)-1)-x_3(1-\eta^2)+(1-\eta)x_B-(\eta^2+\eta-2)z+1)\non
&&-(1-\eta)((1+\eta)(x_B+z)-\eta)] \phi_0(z)-r\sqrt{\eta (1-r^2) }[  (z(1-r^2)+x_B)(\phi_s(z)+\phi_t(z))\non
&&+(1-\eta)x_3(\phi_s(z)-\phi_t(z))+2\phi_s(z)]  \big] E_n(t_v)h_v(x_B,z,x_3,b,b_B)+\big[   (1-\eta +r^2 )\non
&&  \times[(1-r^2)((1-\eta)x_3+\eta z)-x_B\eta ] \phi_0(z)+r\sqrt{\eta (1-r^2) }\big[ (1-\eta)x_3(\phi_s(z)+\phi_t(z))\non
&&+((1-r^2)z-x_B)(\phi_s(z)-\phi_t(z))]\big] E_n(t_w)h_w(x_B,z,x_3,b,b_B)
\big\}.
\eeq

We show here the explicit expressions of the hard functions $h_i$($i=a,b,c,d,e,f,g,h,m,n,o,p,x,y,v,w$), coming from
the Fourier transform of hard kernel:
\beq
h_i(x1,x2(,x3),b_1,b_2)&=&h_1(\beta,b_2)\times h_2(\alpha,b_1,b_2),\nonumber\\
h_1(\beta,b_2)&=&\left\{\begin{array}{ll}
K_0(\sqrt{\beta}b_2), & \quad  \quad \beta >0\\
K_0(i\sqrt{-\beta}b_2),& \quad  \quad \beta<0
\end{array} \right.\nonumber\\
h_2(\alpha,b_1,b_2)&=&\left\{\begin{array}{ll}
\theta(b_2-b_1)I_0(\sqrt{\alpha}b_1)K_0(\sqrt{\alpha}b_2)+(b_1\leftrightarrow b_2), & \quad   \alpha >0\\
\theta(b_2-b_1)I_0(\sqrt{-\alpha}b_1)K_0(i\sqrt{-\alpha}b_2)+(b_1\leftrightarrow b_2),& \quad   \alpha<0
\end{array} \right.
\eeq
where $K_0$, $I_0$ are modified Bessel function with
$K_0(ix)=\frac{\pi}{2}(-N_0(x)+i J_0(x))$ and $J_0$ is the Bessel function,
$\alpha$ and $\beta$ are the factor $i_1, i_2$($i=a,b,c,d,e,f,g,h,m,n,o,p,x,y,v,w$) as defined in the following paragraph.

The hard scale $t_i$ is chosen as the maximum of the virtuality of the internal momentum transition in the hard amplitudes.
For $B_{(s)} \to  \bar{D}_{(s)} \rho \to  \bar{D}_{(s)} \pi \pi$ decays, we have
\beq
t_a&=&\max \{m_B\sqrt{|a_1|},m_B\sqrt{|a_2|}, 1/b, 1/b_B \},~~~~~
t_b=\max \{m_B\sqrt{|b_1|},m_B\sqrt{|b_2|}, 1/b_B, 1/b \};\non
t_c&=&\max \{m_B\sqrt{|c_1|},m_B\sqrt{|c_2|}, 1/b_B, 1/b_3 \},~~~~
t_d=\max \{m_B\sqrt{|d_1|},m_B\sqrt{|d_2|}, 1/b_B, 1/b_3 \};\non
t_e&=&\max \{m_B\sqrt{|e_1|},m_B\sqrt{|e_2|}, 1/b, 1/b_3 \},~~~~~~
t_f=\max \{m_B\sqrt{|f_1|},m_B\sqrt{|f_2|}, 1/b_3, 1/b \};\non
t_g&=&\max \{m_B\sqrt{|g_1|},m_B\sqrt{|g_2|}, 1/b, 1/b_B \},~~~~~
t_h=\max \{m_B\sqrt{|h_1|},m_B\sqrt{|h_2|}, 1/b, 1/b_B \};\non
t_m&=&\max \{m_B\sqrt{|m_1|},m_B\sqrt{|m_2|}, 1/b_3, 1/b_B \},
t_n=\max \{m_B\sqrt{|n_1|},m_B\sqrt{|n_2|}, 1/b_3, 1/b_B \};\non
t_o&=&\max \{m_B\sqrt{|o_1|},m_B\sqrt{|o_2|}, 1/b_B, 1/b \},~~~~~
t_p=\max \{m_B\sqrt{|p_1|},m_B\sqrt{|p_2|}, 1/b_B, 1/b \}.\non
\eeq
with the factors
\beq
a_1&=& (1-r^2 )  z, \qquad a_2= (1-r^2 ) x_B z;\non
b_1&=& (1-r^2 ) (x_B-\eta ), \qquad b_2=a_2;\non
c_1&=&a_2, \qquad c_2= r_c^2-[(1-z)r^2+z][(1-\eta)(1-x_3)-x_B];\non
d_1&=&a_2, \qquad d_2= (1-r^2)z [x_B-(1-\eta ) x_3];\non
e_1&=& r_c^2-[1-z(1-r^2) ], \qquad e_2=(1-r^2)(1-z)[(\eta-1)x_3-\eta];\non
f_1&=& (1-r^2)[(\eta-1)x_3-\eta],\qquad f_2=e_2;\non
g_1&=&e_2, \qquad g_2= 1-[(1-z)r^2+z][(1-\eta)(1-x_3)-x_B];\non
h_1&=&e_2, \qquad h_2= (1-r^2)(1-z)[(\eta-1)x_3-\eta+x_B];\non
m_1&=&(1-\eta) x_3, \qquad m_2=(1-\eta) x_3 x_B;\non
n_1&=&  r_c^2-(r^2-x_B)(1-\eta), \qquad n_2=m_2;\non
o_1&=&m_2, \qquad o_2=[(\eta -1) x_3-\eta ] [(1-z)(1-r^2)-x_B];\non
p_1&=&m_2, \qquad p_2=(1-\eta) x_3  [x_B-(1-r^2)z ].
\eeq

For $B_{(s)} \to  D_{(s)} \rho \to D_{(s)}\pi \pi$ decays, similarly, we have
\beq
t_a&=&\max \{m_B\sqrt{|a_1|},m_B\sqrt{|a_2|}, 1/b, 1/b_B \},~~~~
t_b=\max \{m_B\sqrt{|b_1|},m_B\sqrt{|b_2|}, 1/b_B, 1/b \};\non
t_c&=&\max \{m_B\sqrt{|c_1|},m_B\sqrt{|c_2|}, 1/b_B, 1/b_3 \},~~~
t_d=\max \{m_B\sqrt{|d_1|},m_B\sqrt{|d_2|}, 1/b_B, 1/b_3 \};\non
t_x&=&\max \{m_B\sqrt{|x_1|},m_B\sqrt{|x_2|}, 1/b_3, 1/b \},~~~~
t_y=\max \{m_B\sqrt{|y_1|},m_B\sqrt{|y_2|}, 1/b, 1/b_3 \};\non
t_v&=&\max \{m_B\sqrt{|v_1|},m_B\sqrt{|v_2|}, 1/b, 1/b_B \},~~~~
t_w=\max \{m_B\sqrt{|w_1|},m_B\sqrt{|w_2|}, 1/b, 1/b_B \}.\non
\eeq
with the factors
\beq
a_1&=& (1-r^2 )  z, \qquad a_2= (1-r^2 ) x_B z;\non
b_1&=& (1-r^2 ) (x_B-\eta ), \qquad b_2=a_2;\non
c_1&=&a_2, \qquad c_2= (1-r^2)z (x_B-(1-\eta ) x_3);\non
d_1&=&a_2, \qquad d_2= r_c^2-[(1-z)r^2+z][(1-\eta)(1-x_3)-x_B];\non
x_1&=& (1-r^2)[(\eta-1)x_3-\eta], \qquad x_2=(1-\eta)(r^2-1)x_3z;\non
y_1&=& r_c^2-(1-\eta)[z+r^2(1-z)], \qquad y_2=x_2;\non
v_1&=&x_2, \qquad v_2=[1-(1-\eta)x_3][(1 -r^2)z+x_B]+(1-\eta)x_3 ;\non
w_1&=&x_2, \qquad w_2=(1-\eta)x_3[x_B-(1-r^2)z].
\eeq

%========================= reference=========================%


\begin{thebibliography}{99}

\bibitem{BaBar}
B.~Aubert et al. [BaBar Collaboration], \prd {\bf 78}, 012004 (2008).
%% Evidence for direct CP violation from Dalitz-plot analysis of B \to K\pi\pi

\bibitem{BaBar1}
B.~Aubert et al. [BaBar Collaboration], \prd {\bf 80}, 112001 (2009).
%% Time-dependent amplitude analysis of B \to K_S \pi\pi

\bibitem{BaBar2}
J.~P.~Lees et al. [BaBar Collaboration], \prd {\bf 83}, 112010 (2011).
%% Amplitude analysis of B \to K \pi\pi and evidence of direct CPV in B \to K^* \pi decays

\bibitem{BaBar3}
J.~P.~Lees et al. [BaBar Collaboration], \prd {\bf 85}, 112010 (2012).
%% Study of CPv in Dalitz-plot analyses of B \to 3K

\bibitem{BaBar4}
J.~P.~Lees et al. [BaBar Collaboration], \prd {\bf 85}, 054023 (2012).
%% Amplitude analysis and measurement of the time-dependent CP asymmetry of B \to 3K_S decays

\bibitem{Belle}
P.~Chang et al. [Belle Collaboration], \plb {\bf 599}, 148 (2004).
%% Observation of the decays B \to  K\pi\pi and B \to \rho K

\bibitem{Belle1}
A.~Garmash et al. [Belle Collaboration], \prd {\bf 69}, 012001 (2004).
%% Study of B meson decays to three-body charmless hadronic final states

\bibitem{Belle2}
A.~Garmash et al. [Belle Collaboration], \prd {\bf 71}, 092003 (2005).
%% Dalitz analysis of the three-body charmless decays B \to K \pi\pi  and B \to KKK

\bibitem{Belle3}
A.~Garmash et al. [Belle Collaboration], \prl{\bf 96}, 251803 (2006).
%% Evidence for Large Direct CPV in B \to \rho(770)K^\pm from Analysis of 3B Charmless B \to K\pi\pi Decays

\bibitem{Belle4}
A.~Garmash et al. [Belle Collaboration], \prd {\bf 75}, 012006 (2007).
%% Dalitz analysis of three-body charmless B \to K \pi\pi decay

\bibitem{Belle5}
V.~Gaur et al. [Belle Collaboration], \prd {\bf 87}, 091101 (2013).
%% Evidence for the decay B \to KK\pi

\bibitem{LHCb}
R.~Aaij et al. [LHCb Collaboration], \prl {\bf 111}, 101801 (2013).
%% Measurement of CP Violation in the Phase Space of B \to KK\pi and B \to 3K Decays

\bibitem{LHCb1}
R.~Aaij et al. [LHCb Collaboration], \prl {\bf 112}, 011801 (2014).
%% Measurement of CP Violation in the Phase Space of B \to KK\pi and B \to 3\pi Decays

\bibitem{LHCb2}
R.~Aaij et al. [LHCb Collaboration], \prd {\bf 89}, 092006 (2014).
%% Measurement of resonant and CP components in B^0_s \to Jpsi \pi\pi decays

\bibitem{LHCb3}
R.~Aaij et al. [LHCb Collaboration], \prd {\bf 90}, 012003 (2014).
%% Measurement of the resonant and CP components in B^0\to Jpsi \pi\pi decays

\bibitem{LHCb4}
R.~Aaij et al. [LHCb Collaboration], \prd {\bf 92}, 012012 (2015).
%% Amplitude analysis of B^0 ¡ú \bar{D}^0 K^+\pi^- decays

\bibitem{LHCb5}
R.~Aaij et al. [LHCb Collaboration], \prd {\bf 92}, 032002 (2015).
%% Dalitz plot analysis of B0 ¡ú \bar{D}^0 \pi^+\pi^- decays


\bibitem{Li:2003}
C.H.~Chen and H.n.~Li,  \plb {\bf 561}, 258 (2003).
%% Three-body nonleptonic B decays in perturbative QCD

\bibitem{Li:2004}
C.H.~Chen and H.n.~Li,  \prd {\bf 70}, 054006 (2004).
%% Vector-pseudoscalar two-meson distribution amplitudes in three-body B meson decays

\bibitem{wenfei1}
W.F.~Wang, H.C.~Hu, H.n.~Li and C.D.~L\"u,  \prd {\bf 89}, 074031 (2014).
%% Direct CP asymmetries of three-body B decays in perturbative QCD

\bibitem{wenfei2}
W.F.~Wang, H.n.~Li, W.~Wang and C.D.~L\"u, \prd {\bf 91}, 094024 (2015).
%% S-wave resonance contributions to the B \to Jpsi \pi\pi and Bs \to 2\pi¦Ð2mu decays

\bibitem{Li2}
Y.~Li, A.J.~Ma, W.F.~Wang and Z.J.~Xiao, \epjc {\bf76}, 675 (2016).
%% S-wave resonance contributions to the three-body decays B \to \eta_c(2S)f_0(X) \to \eta_c(2s)\pi\pi in perturbative QCD approach

\bibitem{wenfei}
W.F.~Wang and H.n.~Li, \plb {\bf 763}, 29 (2016).
%% Quasi-two-body decays B \to K\rho \to K\pi\pi in perturbative QCD Approach

\bibitem{li17a}
Y.~Li, A.J.~Ma, W.F.~Wang and Z.J.~Xiao, \prd {\bf 95}, 056008 (2017). %% arXiv:1612.05934.
%% Quasi-two-body decays B \to P\rho \to P\pi\pi in the perturbative QCD approach


\bibitem{Furman}
A.~Furman, R.~Kami\^{n}ski, L.~Le\'{s}niak and B.~Loiseau, \plb  {\bf 622}, 207 (2005).
%% Long-distance effects and final state interactions in B \to \pi\pi K and B \to KKK decays

\bibitem{Furman1}
B.~El-Bennich et al., \prd{\bf 74}, 114009 (2006).
%% Interference between f_0(980) and \rho(770) resonances in B \to \pi\pi K decays

\bibitem{Furman2}
B.~El-Bennich et al., \prd{\bf 79}, 094005 (2009); {\bf 83}, 039903 (2011)(E).
%% CP violation and kaon-pion interactions in B \to K \pi\pi decays

\bibitem{Furman3}
A.~Furman, R.~Kami\^{n}ski, L.~Le\'{s}niak and P. \.Zenczykowski,  \plb  {\bf 699}, 102 (2011).
%% Final state interactions in B \to 3K decays

\bibitem{Cheng}
H.Y.~Cheng, C.K.~Chua, and A.~Soni, \prd  {\bf 72}, 094003 (2005).
%% CP-violating asymmetries in B^0 decays to 3K

\bibitem{Cheng1}
H.Y.~Cheng, C.K.~Chua, and A.~Soni, \prd  {\bf 76}, 094006 (2007).
%% Charmless three-body decays of B mesons

\bibitem{Cheng2}
H.Y.~Cheng, \ijmpa  {\bf 23}, 3229 (2008).
%% Charmless 3-body B Decays: Resonant and Nonresonant Contributions

\bibitem{Cheng3}
H.Y.~Cheng and C.K.~Chua, \prd  {\bf 88}, 114014 (2013).
%% Branching fractions and direct CP violation in charmless three-body decays of B mesons

\bibitem{Cheng4}
H.Y.~Cheng and C.K.~Chua, \prd {\bf 89}, 074025 (2014).
%% Charmless three-body decays of B_s mesons

\bibitem{Cheng5}
H.Y.~Cheng, C.K.~Chua, and Z.Q.~Zhang, \prd  {\bf 94}, 094015 (2016).
%% Direct CP violation in charmless three-body decays of B mesons

\bibitem{Bhattacharya}
B.~Bhattacharya, M.~Gronau and J.~L.~Rosner, \plb {\bf 726}, 337 (2013).
%% CP asymmetries in three-body B¡À decays to charged pions and kaons

\bibitem{Gronau}
M. Gronau, \plb {\bf 727}, 136 (2013).
%% U-spin breaking in CP asymmetries in B decays

\bibitem{Wang}
C.~Wang, Z.H.~Zhang, Z.Y.~Wang and X.H.~Guo, \epjc {\bf 75}, 536 (2015).
%% Localized direct CP violation in B \to \rho(\omega)\pi \to 3\pi

\bibitem{Lesniak}
L.~Les\'niak and P.~\.Zenczykowski, \plb {\bf 737}, 201 (2014).
%% Dalitz-plot dependence of C P asymmetry in B \to 3K decays

\bibitem{LiYing}
Y.~Li, \prd {\bf 89}, 094007 (2014).
%% Comprehensive study of B  \to 2K \pi  decays in the factorization approach

\bibitem{He1}
D.~Xu, G.N.~Li and X.G.~He, \plb {\bf 728}, 579 (2014).
%% U -spin analysis of CPv in B decays into three charged light p mesons

\bibitem{He2}
X.G.~He, G.N.~Li and D.~Xu, \prd {\bf 91}, 014029 (2015).
%% SU(3) and isospin breaking effects on B ¡ú PPP amplitudes

\bibitem{Yang}
Z.H.~Zhang, X.H.~Guo and Y.D.~Yang, \prd  {\bf 87}, 076007 (2013).
%% CPv in B \to 3\pi in the region with low invariant mass of one \pi\pi pair

\bibitem{I}
I.~Bediaga, T.~Frederico, and O.~Louren\c{c}o, \prd {\bf 89}, 094013 (2014).
%% CP violation and CPT invariance in B decays with final state interactions

\bibitem{I1}
J.H.~Alvarenga Nogueira et al., \prd {\bf 92}, 054010 (2015).
%%  CP violation: Dalitz interference, CPT, and final state interactions

\bibitem{Susanne}
S.~Kr\"ankl, T.~Mannel and J.~Virto,\npb {\bf 899}, 247-264 (2015).
%% Three-body non-leptonic B decays and QCD factorization

\bibitem{London}
N.R.-L.~Lorier, M.~Imbeault and D.~London, \prd {\bf 84}, 034040 (2011).
%% Diagrammatic analysis of charmless three-body B decays

\bibitem{London1}
M.~Imbeault, N.R.-L.~Lorier and D.~London, \prd  {\bf 84}, 034041 (2011).
%% Measuring \gamma in B \to K\pi\pi decays

\bibitem{London2}
N.R.-L.~Lorier and D.~London, \prd  {\bf 85}, 016010 (2012).
%% Measuring \gamma  with B \to  K \pi\pi and B \to  KK\bar{K} decays

\bibitem{Beneke}
M.~Beneke, G.~Buchalla, M.~Neubert and C.T.~Sachrajda, \npb {\bf 591}, 313 (2000).
%% QCDF for exclusive non-leptonic B-meson decays: general arguments and the case of heavy¨Clight final states

\bibitem{Bauer}
C.W.~Bauer, D.~Pirjol and I.W.~Stewart, \prl {\bf87}, 201806 (2001).
%%  Proof of Factorization for B \to  D\pi

\bibitem{Chiang}
C.W.~Chiang and E.~Senaha, \prd{\bf75}, 074021 (2007).
%% Updated analysis of two-body charmed B meson decay

\bibitem{sihong}
S.H.~Zhou, Y.B.~Wei, Q.~Qin, Y.~Li, F.S.~Yu and C.D.~L\"u, \prd {\bf 92}, 094016 (2015).
%% Analysis of two-body charmed B meson decays in FATA approach

\bibitem{B-D1}
T.~Kurimoto, H.n.~Li and A.I.~Sanda, \prd {\bf 67}, 054028 (2003).
%% B\to D^(*) form factors in perturbative QCD

\bibitem{B-D11}
Y.Y.~Keum, T.~Kurimoto, H.n.~Li, C.D.~L\"u and A.I.~Sanda, \prd {\bf 69}, 094018 (2004).
%%  Nonfactorizable contributions to B\to D^{(*)} M decays

\bibitem{B-D2}
H.n.~Li, \prd  {\bf 52}, 3958 (1995).
%% Applicability of perturbative QCD to B \to D decays

\bibitem{B-D21}
C.Y.~Wu, T.W.~Yeh and H.n.~Li, \prd  {\bf 53}, 4982 (1996).
%% Perturbative QCD study of B \to D^{(*)} decays

\bibitem{B-D3}
C.D.~L\"u, \epjc {\bf24}, 121 (2002).
%% Calculation of pure annihilation type decay B+ ¡ú Ds +¦Õ

\bibitem{B-D30}
C.D.~L\"u, \prd {\bf 68}, 097502 (2003).
%% Study of color suppressed modes B^0\to \bar{D}^{(*)0} \etap

\bibitem{B-D31}
Y.~Li and C.D.~L\"u, High Energy Phys. $\&$ Nucl. Phys. {\bf 27}, 1062 (2003).
%% B^+ \to D_s^+ \bar{K}^{*0}

\bibitem{Lu1}
R.H.~Li, C.D.~L\"u and Z.~Hao, \prd {\bf 78}, 014018 (2008).
%% B(B_s)\to D_{(s)} (P,V), D_{(s)}^*(P,V) decays in the PQCD approach

\bibitem{Lu11}
Z.~Hao, R.H.~Li, X.X.~Wang and C.D.~L\"u, \jpg {\bf 37}, 015002 (2010).
%% The CKM suppressed B(Bs) \to D_(s)P , ... decays in the perturbative QCD approach

\bibitem{Lu2}
Z.T.~Zou, X.~Yu and  C.D.~L\"u, \prd {\bf 86}, 094001 (2012).
%% B(B_s) \to D_s T, ... decays in perturbative QCD approach

\bibitem{Lu21}
Z.T.~Zou,  R.~Zhou and C.D.~L\"u, \cpc {\bf 37}, 013103 (2013).
%% Pure annihilation type decays B \to D_s K_2^* ... in the perturbative QCD approach

\bibitem{pdg2016}
C.~Patrignani {\it et al.}, (Particle Data Group), \cpc {\bf 40}, 100001 (2016)and  2017 update.

\bibitem{hfag2016}
Y.~Amhis {\it et al}., (Heavy Flavor Averaging Group), arXiv:1612.07233v1 [hep-ex].

\bibitem{Keum:2000wi}
Y.Y.~Keum, H.n.~Li and A.I.~Sanda, \prd {\bf 63}, 054008 (2001).
%% Penguin enhancement and B\to K \pi decays in perturbative QCD

\bibitem{Pire}
M.~Diehl, T.~Gousset, B.~Pire, and O.~Teryaev, \prl {\bf 81}, 1782 (1998).
%% Probing partonic structure in gamma* gamma ---> pi pi near threshold

\bibitem{npb555-231}
M.~V.~Polyakov, Nucl. Phys. B  {\bf 555},  231 (1999).
%%  Hard exclusive electroproduction of two pions and their resonances

\bibitem{JP}
J.~P.~Lees  et al. [BaBar Collaboration], \prd {\bf 86}, 032013 (2012).

 \bibitem{BW}
G.~Breit and E.~Wigner, Phys. Rev. {\bf 49}, 519 (1936).

 \bibitem{GJ}
G.~J.~Gounaris and J.J.~Sakurai, \prl {\bf 21}, 244 (1968).

 \bibitem{Ball}
P.~Ball, G.W.~Jones and R.~Zwicky, \prd {\bf 75}, 054004 (2007).

 \bibitem{Bharucha}
A.~Bharucha, D.M.~Straub and R.~Zwicky, \jhep {\bf 1608}, 098 (2016).

\bibitem{Jansen}
K.~Jansen, C.~McNeile, C.~Michael and C.~Urbach, \prd {\bf 80}, 054510 (2009).

\bibitem{wang}
W.F.~Wang and Z.J.~Xiao, \prd {\bf 86}, 114025 (2012) .


%%%%%%%%%%%%%%%%%%%%%%%%%%%

%------------------------------------------------------
\end{thebibliography}
\end{document}